\documentclass[prb,aps,twocolumn,floats,showpacs,amsmath,amssymb]{revtex4}

\usepackage{graphics}
\usepackage{amsmath}
\usepackage{citesort}
\usepackage{dcolumn}

\begin{document}

\title{
{\rm\small\hfill (submitted to Phys. Rev. B)}\\
  CO oxidation at Pd(100):\\
A first-principles \emph{constrained} thermodynamics study}

\author{Jutta Rogal}
\author{Karsten Reuter}
\author{Matthias Scheffler}
\affiliation{Fritz-Haber-Institut der Max-Planck-Gesellschaft,
Faradayweg 4-6, D-14195 Berlin, Germany}

\received{30th January 2007}

\begin{abstract}
The possible formation of oxides or thin oxide films (surface oxides) on late transition metal surfaces is recently being recognized as an essential ingredient when aiming to understand catalytic oxidation reactions under technologically  relevant gas phase conditions.
Using the CO oxidation at Pd(100) as example, we investigate the composition and structure of this model catalyst surface over a wide range of $(T,p)$-conditions  within a multiscale modeling approach where density-functional theory is linked to thermodynamics.  The results show that under the catalytically most relevant gas phase conditions a thin surface oxide is the  most stable ``phase'' and that the system is actually very close to a transition between this oxidic state and a reduced state in form of a CO covered Pd(100) surface.
\end{abstract}



\maketitle

\section{Introduction}
Catalytic oxidation using transition metals (TM) as the active material is an important technological process, of which we still have only limited microscopic insight.
Much valuable microscopic information that has been obtained under ultra-high vacuum (UHV) conditions is not directly transferable
to the  conditions of practical catalysis (with pressures of several bars and elevated temperatures).
And \emph{in-situ} techniques that would operate under these conditions are still struggling to achieve atomic-scale surface sensitive information.
A key factor that hinders the direct transfer of the UHV insight is that under reaction conditions the entire structure and composition of the catalyst surface might be changed (often called the ``materials gap'').  And  if a new material is created during the induction period of the catalytic process, it will obviously exhibit a very different chemical activity; it will be quantitatively and even qualitatively different.
Under the oxygen-rich gas phase conditions and elevated temperatures of oxidation catalysis the surface of the transition metal catalyst might e.g. be oxidized, so that the actual catalytically active state is not the pristine metal, but rather the formed oxide.
One example for this is the CO oxidation reaction over ruthenium. Here it was found, that under UHV conditions the Ru(0001) model catalyst shows almost no catalytic activity, whereas under high oxygen pressures its catalytic activity exceeds even the one of the frequently used palladium and platinum catalysts~\cite{cant78}. This increase in the catalytic activity was traced back to the formation of a RuO$_2$ film at the surface~\cite{over03,reuter06}. For the ruthenium catalyst a clear distinction could thus be made between the different states, namely the weakly catalytically active  metal under low oxygen pressures and the highly active oxide film under high oxygen pressures.

For the here discussed model system, the CO oxidation at Pd(100), recent \emph{in-situ} experimental measurements also indicate that an oxidic structure at the Pd(100) surface might be formed under the gas phase conditions applied in industrial oxidation catalysis.
Using surface x-ray diffraction (SXRD) measurements a $(T,p)$-diagram showing the detected structures in a pure oxygen atmosphere over a temperature range of $T=600-1000$\,K and pressures of $p_{\mathrm{O}_2} = 10^{-9} - 1$\,atm could be inferred~\cite{lundgren04}.
A thin surface oxide structure was measured over an extended $(T,p)$-range, suggesting that this structure might also appear under catalytic reaction conditions. In reactor scanning tunneling microscopy (STM) experiments by Hendriksen \emph{et al.}~\cite{hendriksen04,hendriksen05} the structure of the Pd(100) surface could directly be monitored during the catalytic oxidation of CO. Here, the surface was exposed to both oxygen and CO at a total pressure of $p_{\mathrm{tot}} \approx 1$\,atm and a temperature of $T \approx 400$\,K. The partial pressures of the reactant gases CO and O$_2$, as well as the reaction product CO$_2$, were measured simultaneously with the STM images. Depending on the partial pressure of O$_2$ and CO the authors observed a change in the reaction rate, which was accompanied by a significant change in the morphology of the surface. This morphology change (albeit not atomically resolved) was interpreted as a change from the adsorbate covered Pd(100) surface to an oxidic state.

To address this problem from a theoretical point of view we employ a multiscale modeling approach where we use density-functional theory (DFT) to describe the system on an electronic (microscopic) level.  By linking these results to thermodynamics~\cite{kaxiras87,scheffler88a,scheffler88b,qian88} it becomes possible to
address much larger system sizes and to
compare the stability of different surface structures in contact with the surrounding gas phase~\cite{wang98,wang00,reuter02,lodzianan03,reuter03a,reuter03b}.
We use in particular  a \emph{constrained} equilibrium approach~\cite{reuter03a,reuter03b}, where the surface is considered to be in full thermodynamic equilibrium with two separate gas phase reservoirs of O$_2$ and CO, and not allowing that O$_2$ and CO can react.
The key result obtained is a surface ``phase diagram'', which provides first insight into  possible surface structures over a wide range of  temperature and pressure conditions of the O$_2$ and CO gas phase.
Focusing on gas phase conditions relevant for oxidation catalysis we find the system  on the verge of either stabilizing a thin surface oxide structure or a CO covered Pd(100) surface.
It is thus well possible and even likely that the surface oxide structure contributes to the active state of the Pd(100) model catalyst under reaction conditions.
However, to verify this, the kinetic effects of the on-going catalytic CO$_2$ formation need to be considered.  This is done in a second step of our hierarchical approach by
performing kinetic Monte Carlo (kMC) simulations on the here identified surface structures, on which we report in a consecutive paper.

\section{Theory}
\subsection{Gibbs free energy of adsorption}
\label{subsec:DeltaG}
To determine the stability of a surface in contact with a gas phase reservoir we use the surface free energy $\gamma$.
For a multi-component system in equilibrium with atomic reservoirs (e.g. a surrounding gas or liquid phase environment, or a macroscopic bulk phase) a general expression for the surface free energy is given by
\begin{eqnarray}
\gamma(T,\{p_i\}) = \frac{1}{A}\left[G^{\mathrm{surf}} - \sum_i N_i \mu_i(T,p_i)\right] \quad .
\end{eqnarray}
Here, $G^{\mathrm{surf}}$ is the Gibbs free energy of the solid including the surface,  $A$ is the surface area and $\mu_i(T,p_i)$ are the chemical potentials of the various species $i$ in the system.  To compare the stability of different adsorption structures of oxygen and CO on the Pd(100) surface depending on the surrounding gas phase conditions one has to evaluate the Gibbs free energy of adsorption, $\Delta G^{\mathrm{ads}}$.~\cite{li03,reuter04a} For this, the stability of the different adsorption structures is compared with respect to the clean metal surface. For a Pd(100) surface in a \emph{constrained equilibrium} with an O$_2$ and CO gas phase $\Delta G^{\mathrm{ads}}$ is thus given by
\begin{eqnarray}
\label{eq:DeltaG}
\lefteqn{\Delta G^{\mathrm{ads}} (\Delta \mu_{\mathrm{O}},\Delta \mu_{\mathrm{CO}}) =} \\ \nonumber
&=& \gamma_{\mathrm{Pd(100)}}-\gamma_{\mathrm{O,CO@Pd(100)}}\\ \nonumber
&=&  -\frac{1}{A}\Big( G^{\mathrm{surf}}_{\mathrm{O,CO@Pd(100)}} - G^{\mathrm{surf}}_{\mathrm{Pd(100)}}
- \Delta N_{\mathrm{Pd}} \mu_{\mathrm{Pd}}^{\mathrm{bulk}} \\ \nonumber
&&- N_{\mathrm{O}}(1/2 E^{\mathrm{tot}}_{\mathrm{O}_2} + \Delta \mu_{\mathrm{O}})
- N_{\mathrm{CO}}(E^{\mathrm{tot}}_{\mathrm{CO}} + \Delta \mu_{\mathrm{CO}})\Big) \\ \nonumber
&\approx&  -\frac{1}{A}\widetilde{E}^{\mathrm{bind}}_{\mathrm{O,CO@Pd(100)}}
+ \frac{N_{\mathrm{O}}}{A} \Delta \mu_{\mathrm{O}}
+ \frac{N_{\mathrm{CO}}}{A} \Delta \mu_{\mathrm{CO}}\quad ,
\end{eqnarray}
where $G^{\mathrm{surf}}_{\mathrm{O,CO@Pd(100)}}$ is the Gibbs free energy of the Pd(100) surface with $N_{\mathrm{O}}$ adsorbed O atoms and $N_{\mathrm{CO}}$ adsorbed CO molecules, and $G^{\mathrm{surf}}_{\mathrm{Pd(100)}}$ is the Gibbs free energy of the clean palladium surface. If the number of palladium atoms per surface area $A$ in the adsorption structure, $N_{\mathrm{Pd}}$,  and in the clean surface, $N'_{\mathrm{Pd}}$, are not equal (e.g. due to a surface reconstruction), i.e. $\Delta N_{\mathrm{Pd}} = N_{\mathrm{Pd}} - N'_{\mathrm{Pd}} \neq 0$, then the excess/deficiency atoms are taken from/put into a bulk reservoir, represented by the chemical potential of the bulk solid phase, $\mu^{\mathrm{bulk}}_{\mathrm{Pd}}$.
The chemical potentials of the two gas phase reservoirs of oxygen, $\mu_{\mathrm{O}_2}$, and CO, $\mu_{\mathrm{CO}}$, have been separated into a total energy contribution and the remaining part containing all the temperature and pressure dependent terms, i.e. $\mu_{\mathrm{O}} = 1/2 \mu_{\mathrm{O}_2} = 1/2 E^{\mathrm{tot}}_{\mathrm{O}_2} + \Delta \mu_{\mathrm{O}} (T,p)$, setting $\Delta \mu_{\mathrm{O}} (T,p) = 1/2 \Delta \mu_{\mathrm{O}_2} (T,p)$, and $\mu_{\mathrm{CO}} = E^{\mathrm{tot}}_{\mathrm{CO}} + \Delta \mu_{\mathrm{CO}} (T,p)$~\cite{reuter02,reuter03b}.
In the last line of Eq.~(\ref{eq:DeltaG}) the difference of the Gibbs free energies of the clean and adsorbate covered surfaces, and of the bulk system, has been approximated by the DFT total energies $E$~\cite{reuter02,reuter03b}.  A detailed discussion of the thereby neglected contributions to the Gibbs free energy of adsorption and their influence on the presented results will be given below.

In the constrained equilibrium approach leading to Eq.~(\ref{eq:DeltaG}) the two gas phases are assumed to be non-interacting~\cite{reuter03b}.  The Gibbs free energy of adsorption  $\Delta G^{\mathrm{ads}}$ depends then linearly on the chemical potentials of both the oxygen and CO gas phase.  The slope is respectively determined by the surface coverage of oxygen and/or CO.
This is most apparent in the last line of Eq.~(\ref{eq:DeltaG}), where we have introduced the average binding energy of O and CO as
%
\begin{eqnarray}
\label{eq:Ebind}
\lefteqn{\widetilde{E}^{\mathrm{bind}}_{\mathrm{O,CO@Pd(100)}}=} \\ \nonumber
&=& \Big(E^{\mathrm{tot}}_{\mathrm{O,CO@Pd(100)}} - E^{\mathrm{tot}}_{\mathrm{Pd(100)}}
- \Delta N_{\mathrm{Pd}} E^{\mathrm{tot}}_{\mathrm{Pd}} \\ \nonumber
&  & - \frac{N_{\mathrm{O}}}{2} E^{\mathrm{tot}}_{\mathrm{O}_2}
- N_{\mathrm{CO}} E^{\mathrm{tot}}_{\mathrm{CO}}\Big)
\quad .
\end{eqnarray}
%
Since the most stable structure will be the one with the lowest surface free energy, an adsorbate structure will be stable with respect to the clean surface, if $\gamma_{\mathrm{O,CO@Pd(100)}} < \gamma_{\mathrm{Pd(100)}}$, i.e. if $\Delta G^{\mathrm{ads}} > 0$.
If two structures contain  an equivalent amount of oxygen and CO they will also show the same dependence on $\Delta \mu_{\mathrm{O}}$ and $\Delta \mu_{\mathrm{CO}}$.  The ratio of how these two structures contribute to the stable structures is governed by the law of mass action, which enables us to directly exclude a less favorable structure, if its binding energy, $\widetilde{E}^{\mathrm{bind}}_{\mathrm{O,CO@Pd(100)}}$, differs by much more than $k_{\mathrm{B}}T$ from the one of the most favorable structure.

For the adsorption of O/CO on the Pd(100) surface the average binding energy defined in Eq.~(\ref{eq:Ebind}) is equivalent to the commonly used binding energy, which is often given per O atom respectively CO molecule, e.g. in the case of oxygen as
\begin{eqnarray}
\label{eq:Ebind_O}
\lefteqn{E^{\mathrm{bind}}_{\mathrm{O@Pd(100)}}=} \\ \nonumber
&=& \frac{1}{N_{\mathrm{O}}}\left(E^{\mathrm{tot}}_{\mathrm{O@Pd(100)}} - E^{\mathrm{tot}}_{\mathrm{Pd(100)}}
- \frac{N_{\mathrm{O}}}{2} E^{\mathrm{tot}}_{\mathrm{O}_2}\right)
\quad .
\end{eqnarray}
In the case of additional adsorption of O/CO on the reconstructed $\sqrt{5}$ surface oxide (which will be introduced below) the binding energy of the atom/molecule with respect to this $\sqrt{5}$ structure is then given by (here exemplified for CO)
\begin{eqnarray}
\label{eq:Ebind_sqrt5}
\lefteqn{E^{\mathrm{bind}}_{\mathrm{CO@}\sqrt{5}}=} \\ \nonumber
& = &  \frac{1}{N_{\mathrm{CO}}}\left(E^{\mathrm{tot}}_{\mathrm{CO@}\sqrt{5}} - E^{\mathrm{tot}}_{\sqrt{5}}
- N_{\mathrm{CO}} E^{\mathrm{tot}}_{\mathrm{CO}}\right)
\quad ,
\end{eqnarray}
where $E^{\mathrm{tot}}_{\sqrt{5}}$ is the total energy of the $\sqrt{5}$ surface oxide.
We note that $E^{\mathrm{bind}}$ is in this case different to the above defined $\widetilde{E}^{\mathrm{bind}}$ in Eq.~(\ref{eq:Ebind}), since the latter also contains the energetic changes due to the formation of the $\sqrt{5}$ surface oxide, i.e. it is referenced with respect to the clean Pd(100) surface.

\subsection{Bulk oxide stability}
\label{subsec:bulkoxide}
When considering a metal surface in contact with an O$_2$ and CO gas phase
complete conversion into an extended bulk oxide is also a thermodynamic possibility.
The stability of the corresponding bulk oxide has therefore to be evaluated with respect to the two gas phase components and compared to the various stable structures forming at the metal surface~\cite{reuter04a}.  A sufficient oxygen content in the gas phase will lead to the formation of bulk palladium oxide, PdO, whereas a sufficient CO content will favor the decomposition of PdO into CO$_2$ and Pd metal.
In a pure oxygen gas phase the thermodynamic stability of the bulk oxide is given by
\begin{eqnarray}
\label{eq:bulkoxide1}
\mu^{\mathrm{bulk}}_{\mathrm{PdO}} < \mu^{\mathrm{bulk}}_{\mathrm{Pd}} + \mu_{\mathrm{O}} \quad ,
\end{eqnarray}
whereas in a pure CO environment the stability condition for the oxide is
\begin{eqnarray}
\label{eq:bulkoxide2}
\mu^{\mathrm{bulk}}_{\mathrm{PdO}} + \mu_{\mathrm{CO}} < \mu^{\mathrm{bulk}}_{\mathrm{Pd}} + \mu_{\mathrm{CO}_2} \quad .
\end{eqnarray}
If we approximate the chemical potential of the free CO$_2$ molecule by its total energy only~\cite{reuter04a}, Eq.~(\ref{eq:bulkoxide1}) and Eq.~(\ref{eq:bulkoxide2}) can be combined yielding the stability criterion for PdO in an oxygen and CO containing gas phase
\begin{eqnarray}
\label{eq:bulkoxide3}
\Delta \mu_{\mathrm{CO}} - \Delta \mu_{\mathrm{O}} \lesssim - 2 \Delta H^f_{\mathrm{PdO}}(T=0\,\mathrm{K}) + \Delta E^{\mathrm{mol}} \quad .
\end{eqnarray}
Here, $\Delta H^f_{\mathrm{PdO}} \approx E^{\mathrm{tot}}_{\mathrm{PdO}} - E^{\mathrm{tot}}_{\mathrm{Pd}} - 1/2 E^{\mathrm{tot}}_{\mathrm{O}_2}$ is the heat of formation of PdO at $T = 0$\,K, and $\Delta E^{\mathrm{mol}} = E^{\mathrm{bind}}_{\mathrm{CO}_2} - E^{\mathrm{bind}}_{\mathrm{CO}} - 1/2 E^{\mathrm{bind}}_{\mathrm{O}_2}$ is the difference in binding energies between the three gas phase molecules.
Table~\ref{tab:Eb_gas} compiles these binding energies, as well as the heat of formation of PdO, as obtained by previously published highly converged DFT calculations~\cite{kiejna06} and using the computational methodology described below.
In addition the experimental values are quoted.  These values have been extrapolated to $T=0$\,K and the zero point vibration energy (ZPE, also listed in Table~\ref{tab:Eb_gas}) has been removed.

Substituting the chemical potential of CO$_2$, $\mu_{\mathrm{CO}_2}$, by only the total energy term, $E^{\mathrm{tot}}_{\mathrm{CO}_2}$, in Eq.~(\ref{eq:bulkoxide2}) is a rather crude approximation.  The most important contributions that have been neglected arise from the vibrational and translational free energy.  The assumption that the CO$_2$ formed at the surface is readily transported away, motivates us to disregard the translational free energy contribution~\cite{reuter04a}.
The vibrational free energy contribution will be of the order of the ZPE for the temperature range discussed here.  As can be seen in Table~\ref{tab:Eb_gas} the ZPE value for CO$_2$ is approximately 0.3\,eV.
Variations of $\Delta E^{\mathrm{mol}}$ in Eq.~(\ref{eq:bulkoxide3}) of this order of magnitude do not affect any of the conclusions discussed below, which justifies the rather crude approximation for the present purpose.
\begin{table}
\caption{\label{tab:Eb_gas}
Computed heat of formation $\Delta H^f_{\mathrm{PdO}} (0,0)$ and binding energies of O$_2$, CO and CO$_2$ (see text).  The quoted experimental values are extrapolated to $T=0$\,K and the zero point vibration energy (ZPE) has been removed. All values are in eV.}
\begin{ruledtabular}
\begin{tabular}{lrrrrl}
\rule[0mm]{0mm}{4mm}         & PBE &   RPBE    &   LDA &   Exp.~\cite{handbook} & ZPE   \\[0.5ex]
\hline
\rule[0mm]{0mm}{4mm}$\Delta H^f_{\mathrm{PdO}} (0,0)$    &   -0.87   &   -0.62   &  -1.42  &  -0.97\\
\\
$E^{\mathrm{bind}}_{\mathrm{O}_2}$    & -6.22   &   -5.75   &   -7.56   &   -5.30 & (0.10)  \\[0.1ex]
$E^{\mathrm{bind}}_{\mathrm{CO}}$    & -11.65  &   -11.20  &   -12.93  &   -11.33 & (0.13) \\[0.1ex]
$E^{\mathrm{bind}}_{\mathrm{CO}_2}$  & -17.99  &   -17.09  &   -20.43  &   -16.98 & (0.27)  \\[0.5ex]
\hline
\rule[0mm]{0mm}{4mm}$\Delta E^{\mathrm{mol}}$      &   -3.24   &   -3.02   &   -3.72   &   -3.00\\[0.5ex]
\end{tabular}
\end{ruledtabular}
\end{table}

\subsection{Computational Setup}
With the approximations made in the last two subsections, the crucial quantities determining the stability of surface structures are the total energies of the extended surfaces and of the involved gas phase molecules.  These total energies are obtained by DFT calculations which have been performed
within the full-potential (linearized) augmented plane wave + local orbital (L)APW+lo method~\cite{sjoestedt00,madsen01} as implemented in the \textsf{WIEN2k} code~\cite{wien2k}.

All surfaces are simulated within the supercell approach, using inversion symmetric slabs consisting of five Pd(100) layers with adsorption of oxygen and/or CO or the reconstructed surface oxide plus additional O/CO on both sides.  The vacuum between consecutive slabs is at least 14\,\AA.  The adsorption layers and two outermost palladium layers have been fully relaxed.
The muffin-tin radii have been set to $R^{\mathrm{Pd}}_{\mathrm{MT}}=2.0$\,bohr for palladium,
$R^{\mathrm{O}}_{\mathrm{MT}}=1.0$\,bohr for oxygen and $R^{\mathrm{C}}_{\mathrm{MT}}=1.0$\,bohr for carbon.
Inside the muffin-tins the wave functions are expanded up to $l^{\mathrm{wf}}_{\mathrm{max}} = 12$ and the potential up to $l^{\mathrm{pot}}_{\mathrm{max}} = 6$.
For the Pd(100)-$(1 \times 1)$ structure a $[10 \times 10 \times 1]$ Monkhorst-Pack (MP) grid has been used to integrate the Brillouin zone (BZ).  For larger surface unit cells the MP-grid has been reduced accordingly to assure an equivalent sampling of the BZ.  Since calculations with different MP-grids are not fully comparable, the energies of the clean metal surface, $E^{\mathrm{tot}}_{\mathrm{Pd(100)}}$, and of bulk palladium, $E_{\mathrm{Pd}}^{\mathrm{tot}}$, needed to evaluate the Gibbs free energy of adsorption as defined in Eq.~(\ref{eq:DeltaG}), have been calculated each time in the same supercell as the corresponding adlayers or surface oxide structure.
The energy cutoff for the expansion of the wave function in the interstitial is $E^{\mathrm{wf}}_{\mathrm{max}} = 20$\,Ry and for the potential $E^{\mathrm{pot}}_{\mathrm{max}} = 196$\,Ry.  Using these basis set parameters the average binding energies per oxygen atom/CO molecule are converged within 50\,meV, so that the Gibbs free energies of adsorption are converged within 1–-5\,meV/\AA$^2$.

All molecular calculations are done in rectangular supercells with side lengths of $(13 \times 14 \times 15)$\,bohr (atoms) or  $(13 \times 14 \times 20)$\,bohr (molecules) using only the $\Gamma$-point to sample the BZ.
To obtain the binding energies of the three gas phase molecules, O$_2$, CO and CO$_2$, very accurately the cutoff for the expansion of the wave function in the interstitial has been increased to $E^{\mathrm{wf}}_{\mathrm{max}} = 37$\,Ry.  With respect to this cutoff the binding energies are converged within 10-20\,meV.

The exchange-correlation (xc) energy is treated within the generalized gradient approximation (GGA) using the PBE~\cite{perdew96} xc-functional.  To assess the error introduced to the total energies by the  approximate xc-energy, the most important calculations have also been repeated using another gradient corrected xc-functional, the RPBE~\cite{hammer99}, as well as the local-density approximation, LDA~\cite{perdew92}.  The influence of these different xc-functionals on the results will be discussed below.

\section{O and/or CO adsorption at P\lowercase{d}(100)}
In this section we discuss the different adlayer structures of oxygen and/or CO on Pd(100) that will be compared within the constrained thermodynamic equilibrium approach.  In addition to the  on-surface adsorption of O/CO on Pd(100),  the formation of a reconstructed $(\sqrt{5} \times \sqrt{5})R27^{\circ}$ (abbreviated with $\sqrt{5}$ in the following) surface oxide  has experimentally been observed at higher oxygen exposure~\cite{orent82,chang88a,chang88b,zheng02}. Since the structure of the $\sqrt{5}$ surface oxide is quite different from the clean Pd(100) surface, also exhibiting different adsorption sites, we will discuss the two surfaces separately.

\subsection{Adlayers on Pd(100)}
\subsubsection{Pure oxygen adsorption}
Experimentally, four different adsorption structures have been observed when exposing the Pd(100) surface to oxygen~\cite{orent82,stuve84,chang88a,chang88b,zheng02}.  These are a $p(2 \times 2)$, a $c(2 \times2)$, a $(5 \times 5)$ and the $\sqrt{5}$ surface oxide structure.
The $\sqrt{5}$  surface oxide will be discussed in detail in the next section.
The $(5 \times 5)$ structure will not been considered in this study, since the structure and state of the oxygen atoms has not been well established so far, i.e. there exists no well defined structural model. In addition, the $(5 \times 5)$ structure appears to be only of metastable character and its formation was found to be very sensitive to the surface preparation and oxygen exposure range, so that in the oxidation as well as in the reduction process the $(5 \times 5)$ structure can be bypassed going directly from a $(2 \times 2)$ to a $\sqrt{5}$ structure and vice versa~\cite{zheng02}.

This leaves as  ordered adlayers only the $p(2 \times 2)$ at 0.25 monolayer (ML) coverage and the $c(2 \times2)$ at 0.5\,ML.
Computing the binding energy of O atoms in all high-symmetry sites offered by the (100) surface (bridge, top and hollow), we find
in both structures the fourfold hollow site to be the most stable adsorption site.
The binding energies are calculated with respect to the O$_2$ gas phase molecule (cf. Eq.~(\ref{eq:Ebind_O}))
and the values of the binding energies per oxygen atom are listed in Table~\ref{tab:Eb_simple}.  It can be seen that the binding energy decreases going from the $p(2 \times 2)$ to the higher coverage $c(2 \times 2)$ adlayer, indicating overall repulsive interactions between the adsorbed oxygen atoms.

\subsubsection{Pure CO adsorption}
\begin{figure}
\scalebox{0.50}{\includegraphics{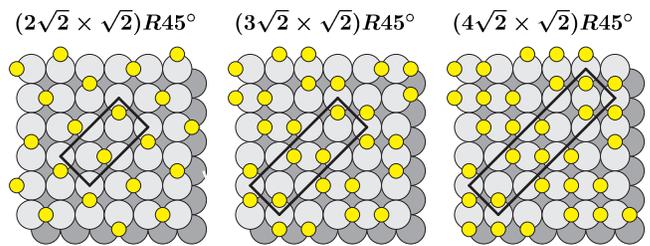}}
\caption{\label{fig1}
(Color online) Top views of the three experimentally characterized adlayer structures of CO on Pd(100). From left to right the coverage increases from $\Theta = 0.5$ to $\Theta = 0.67$ and $\Theta = 0.75$\,ML.  The yellow small spheres represent the CO molecules, the grey large spheres the Pd(100) surface (second layer atoms are darkened).}
\end{figure}

The adsorption of CO on Pd(100) has been studied intensively~\cite{bradshaw78,hoffmann83,behm79,behm80,uvdal88,yoshioka90,andersen91,berndt92}.  In all experimental studies it is found that  CO binds upright in the bridge position via the C atom.  At a coverage of $\Theta = 0.50$\,ML an ordered $(2 \sqrt{2} \times \sqrt{2})R45^{\circ}$ adlayer is formed, which is compressed to a $(3 \sqrt{2} \times \sqrt{2})R45^{\circ}$ structure at $\Theta = 0.67$\,ML and a $(4 \sqrt{2} \times \sqrt{2})R45^{\circ}$ at $\Theta = 0.75$\,ML (cf. Fig.~\ref{fig1}).
The formation of the $(2 \sqrt{2} \times \sqrt{2})R45^{\circ}$ adlayer rather than a simple $c(2 \times 2)$ structure is assumed to be mainly due to  overall strongly repulsive interactions among the adsorbed CO molecules~\cite{eichler98}.  In the $c(2 \times 2)$ each CO molecule has four nearest neighbors at a distance of $\sqrt{2} a$, where $a$ is the length of the $(1 \times 1)$ surface unit cell.  In the $(2 \sqrt{2} \times \sqrt{2})R45^{\circ}$ structure, however, the CO molecules form a distorted hexagonal overlayer with only two nearest neighbors at a distance of $\sqrt{2} a$ and four neighbors slightly further away at a distance of $\sqrt{5/2} a$.

In addition to these experimentally observed structures also simple $(1 \times 1)$ and $(2 \times 2)$ adlayers with CO in top, bridge and hollow sites are considered in this work.  The adsorption in bridge sites is found to be  energetically favored at all computed coverages.  Table~\ref{tab:Eb_simple} compiles therefore only the results for bridge-bonded CO.  Looking at the values in Table~\ref{tab:Eb_simple} it can be seen that the binding energy per CO molecule is almost constant up to a coverage of $\Theta = 0.50$\,ML.  For higher coverages the binding energy decreases due to overall repulsive interactions between the adsorbed CO molecules.  Comparing the two discussed structures with a coverage of $\Theta = 0.50$\,ML we in fact find
them to be degenerate within our computational accuracy (the $(2 \sqrt{2} \times \sqrt{2})R45^{\circ}$ is only favored by a few meV). These results are also fully consistent with previous DFT pseudopotential studies~\cite{hammer99,eichler98}.

\begin{table}
\caption{\label{tab:Eb_simple}
Calculated binding energies of on-surface adlayers of O/CO on Pd(100).  For the pure adlayer structures the binding energies are given per O atom resp. CO molecule.  For the mixed structures the total binding energies are given.  All values are in eV.}
\begin{ruledtabular}
\begin{tabular}{lrr}
\\ [-0.2cm]
    &  Coverage $\Theta$ &  $E^{\mathrm{bind}}$\\ [0.1cm]\hline
\\[-0.1cm]
\textbf{Pure O structures:}\\[0.1cm]
$p(2 \times 2)$-O$^{\mathrm{hol}}$   &  0.25     &  -1.35 \\
$c(2 \times 2)$-O$^{\mathrm{hol}}$   &  0.50     &  -1.10 \\
\\
\textbf{Pure CO structures:}\\[0.1cm]
$p(2 \times 2)$-CO$^{\mathrm{br}}$   &  0.25     &  -1.92 \\
$c(2 \times 2)$-CO$^{\mathrm{br}}$   &  0.50     &  -1.92 \\
$(2 \sqrt{2} \times \sqrt{2})R45^{\circ}$-CO$^{\mathrm{br}}$   &  0.50 &  -1.92\\
$(3 \sqrt{2} \times \sqrt{2})R45^{\circ}$-CO$^{\mathrm{br}}$   &  0.67 &  -1.73\\
$(4 \sqrt{2} \times \sqrt{2})R45^{\circ}$-CO$^{\mathrm{br}}$   &  0.75 &  -1.63\\
$(1 \times 1)$-CO$^{\mathrm{br}}$   &  1.00     &  -1.31 \\
\\
\textbf{Mixed O/CO structures:}\\[0.1cm]
$(2 \times 2)$-O$^{\mathrm{hol}}$-CO$^{\mathrm{br}}$   &  0.50     &  -2.95 \\
$(2 \times 2)$-2O$^{\mathrm{hol}}$-CO$^{\mathrm{br}}$   &  0.75     &  -2.93 \\
$(2 \times 2)$-O$^{\mathrm{hol}}$-2CO$^{\mathrm{br}}$   &  0.75     &  -3.96 \\
$(2 \times 2)$-2O$^{\mathrm{hol}}$-2CO$^{\mathrm{br}}$   &  1.00     &  -3.64 \\
\end{tabular}
\end{ruledtabular}
\end{table}

\subsubsection{Mixed O/CO adsorption}
As to  the adsorption of oxygen and CO on the Pd(100) surface, experimentally no ordered overlayers have been reported so far.  If the Pd(100) surface is exposed to both oxygen and CO, the two adsorbed species  tend to form separate domains instead~\cite{stuve84}.  In their low energy electron diffraction (LEED) measurements Stuve \textit{et al.}~\cite{stuve84} observed that for a fully developed $p(2 \times 2)$-O/Pd(100) structure the $p(2 \times 2)$ LEED pattern vanished after exposure to CO at a temperature of $T=80$\,K.  Assuming a barrier of $\approx 1.0$\,eV for the reaction of O and CO to form CO$_2$~\cite{zhang01,eichler02} it is rather unlikely that the adsorbed oxygen reacted with the CO at this low temperature.  The disappearance of the LEED signal was thus interpreted as a CO induced disordering of the oxygen islands.
\begin{figure}
\scalebox{0.5}{\includegraphics{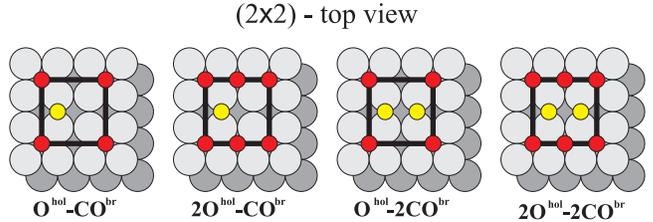}}
\caption{\label{fig2}
(Color online) Top views of the considered co-adsorbed overlayer structures of oxygen and CO on Pd(100). The small red (dark) spheres correspond to oxygen atoms, the small yellow (light) spheres represent the CO molecules and the large grey spheres the Pd(100) surface (second layer atoms are darkened).}
\end{figure}
To nevertheless obtain an idea about the simultaneous adsorption of O and CO on Pd(100) we set up different models for  mixed overlayer structures that seemed to be most obvious from a combinatorial point of view.  For this we used $(2 \times 2)$ surface unit cells placing 1 to 2 O atoms and CO molecules in their favorite adsorption sites, i.e. O in hollow and CO in bridge sites.  Excluding any structures where O and CO are closer to each other than the length $a$ of the $(1 \times 1)$ surface unit cell we obtain the four different structures shown in Fig.~\ref{fig2}.
The corresponding  average binding energies calculated using Eq.~(\ref{eq:Ebind})  are listed in Table~\ref{tab:Eb_simple}.

Similar to the pure adlayer structures of O or CO on Pd(100), the binding energy per adsorbate decreases with increasing coverage in the considered co-adsorption structures.  This trend in binding energies reflects the afore mentioned repulsive interactions among the adsorbed O atoms and CO molecules.  For a coverage of $\Theta = 0.50$\,ML the binding energy of the mixed O/CO overlayer is   less favorable by $70$\,meV than the sum of the binding energies of the respective $c(2 \times 2)$ pure adlayers of O and CO. This indicates that the repulsive interactions among adsorbed O and CO are even stronger than between O and O, respectively CO and CO in the pure adlayer structures.  The adsorption of both O and CO in hollow sites in a $c(2 \times 2)$ structure further decreases the binding strength of the adsorbates.

The even stronger repulsive interactions in these mixed adsorbate structures will thus favor the separation of oxygen and CO into separate domains. The adsorption of CO into a $p(2 \times 2)$-O structure can then induce a demixing of the two species and destroy the long-range order, which is consistent with the disappearance of the LEED spot intensities in the afore mentioned experiment by Stuve \emph{et al.}~\cite{stuve84}.

\subsection{O and CO on the surface oxide}
\begin{figure}
\scalebox{0.37}{\includegraphics{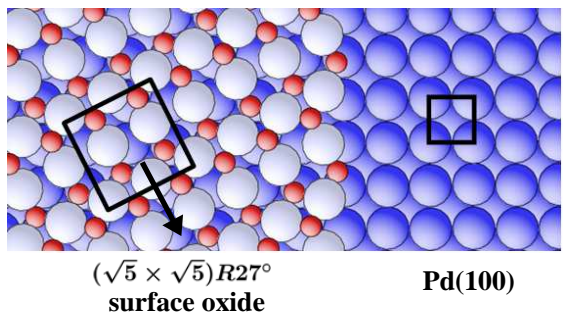}}
\caption{\label{fig3a}
(Color online)  Top view of the $(\sqrt{5} \times \sqrt{5})R27^{\circ}$ surface oxide structure (left part) and the Pd(100) substrate (right part) and their corresponding surface unit cells (black lines). Red small spheres represent the lattice oxygen atoms, large light-blue ones the reconstructed Pd atoms and large dark-blue spheres Pd atoms belong to the underlying Pd(100) substrate.  The arrow indicates the easy shift direction of the surface oxide trilayer along the Pd(100) substrate.}
\end{figure}
At a coverage of $\Theta = 0.80$\,ML, the adsorbed oxygen induces a reconstruction of the Pd(100) surface forming  the $\sqrt{5}$ surface oxide structure.  An atomistic model for the surface oxide (cf. Fig.~\ref{fig3a}) was proposed in a detailed experimental and theoretical study, revealing that the structure can actually be described as a PdO(101) trilayer on Pd(100)~\cite{todorova03}.  The surface unit cell contains four palladium and four oxygen atoms.  Two Pd atoms are fourfold  and the other two are twofold coordinated by oxygen.  There are also two kinds of O atoms. Two O atoms sit on-top of the reconstructed Pd layer and two at the interface to the Pd(100) substrate forming the afore mentioned trilayer structure.

In addition to the previous discussion on this surface oxide structure~\cite{todorova03}, we note that the potential energy surface for a registry shift of the entire surface oxide trilayer over the Pd(100) substrate is very shallow.  In particular, shifts of up to $\sim 0.5$\,\AA \, along the direction as indicated by the arrow in Fig.~\ref{fig3a} and parallel to the surface lead only to energy variations of less than 50\,meV per $\sqrt{5}$ surface unit cell.
Without anchoring by e.g. defects, the lateral position of the ideal surface oxide over the Pd(100) surface is thus not well defined.  In extended test calculations, we verified that this uncertainty in the lateral position has no consequences on the adsorption energetics discussed here and therefore simply use in the present work the lateral position determined in Ref.~\onlinecite{todorova03}.

\begin{figure}
\scalebox{0.42}{\includegraphics{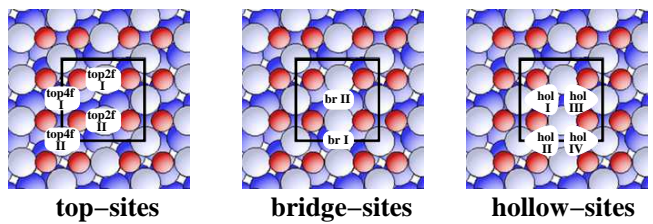}}
\caption{\label{fig3}
(Color online) Top views of high-symmetry adsorption sites on the $(\sqrt{5} \times \sqrt{5})R27^{\circ}$ surface oxide structure. Red small spheres represent the lattice oxygen atoms, large light-blue ones the reconstructed Pd atoms and large dark-blue spheres Pd atoms belong to the underlying Pd(100) substrate.}
\end{figure}

For the additional adsorption of oxygen and/or CO on the surface oxide structure there are several possibilities.  As can be seen in Fig.~\ref{fig3} the surface oxide exhibits top, bridge and hollow high-symmetry sites.
Due to the symmetry of the underlying Pd(100) substrate multiple adsorption sites of the same type (top, bridge or hollow) are not fully equivalent, but still very similar.  Only the top sites differ substantially depending on whether the corresponding Pd atom is twofold or fourfold coordinated by oxygen. As yet there is no  experimental information available for the additional adsorption of O or CO on the $\sqrt{5}$ surface oxide.  Therefore, we performed a systematic investigation of possible overlayer structures of O and CO involving the reconstructed $\sqrt{5}$ surface oxide and starting with the adsorption sites depicted in Fig.~\ref{fig3}.

In a first step only one CO molecule is adsorbed in any of the ten adsorption sites.  We find that the four different hollow sites are not stable, i.e.  upon relaxation the adsorbed CO molecule always moves to the respective neighboring bridge site.  The binding energies of the two bridge (br~I and br~II) sites are almost degenerate as expected from the only slight structural differences, i.e. the binding energies as defined in Eq.~(\ref{eq:Ebind_sqrt5}) differ by less than 0.1\,eV.  Similarly, the two top sites at the two twofold (top2f~I and top2f~II)  oxygen coordinated Pd atoms exhibit nearly equivalent binding energies ($\Delta E^{\mathrm{bind}} < 50$\,meV).  The same is also found for the two top sites at the two fourfold (top4f~I and top4f~II)  oxygen coordinated Pd atoms.
In the following we will thus treat these correspondingly very similar sites as being identical.
The bridge site is the most stable adsorption site followed by the top2f site, and on the top4f site the CO molecule is already only weakly bound (cf. Table~\ref{tab:Eb_sqrt5}).
\begin{small}
\begin{table}
\caption{\label{tab:Eb_sqrt5}Calculated binding energies of CO on the $\sqrt{5}$ surface oxide structure calculated using Eq.~(\ref{eq:Ebind_sqrt5}).  2CO denotes that two CO molecules are adsorbed in the $\sqrt{5}$ surface unit cell.  The adsorption of two CO molecules in top4f sites or one CO in bridge and a second in a top2f site does not lead to a stable structure. All values are per CO molecule and in eV.}
\begin{ruledtabular}
\begin{tabular}{llllll}
\rule[0mm]{0mm}{4mm}     & $E^{\mathrm{bind}}_{\mathrm{CO@}\sqrt{5}}$ &      & $E^{\mathrm{bind}}_{\mathrm{CO@}\sqrt{5}}$  &      & $E^{\mathrm{bind}}_{\mathrm{CO@}\sqrt{5}}$  \\[0.5ex]
\hline
\rule[0mm]{0mm}{4mm}CO$^{\mathrm{br}}$  & -0.93 & 2CO$^{\mathrm{br}}$    & -0.75 & 2CO$^{\mathrm{br,top2f}}$ & --- \\
CO$^{\mathrm{top2f}}$                   & -0.62 & 2CO$^{\mathrm{top2f}}$ & -0.51 & 2CO$^{\mathrm{br,top4f}}$ & -0.45  \\
CO$^{\mathrm{top4f}}$                   & -0.13 & 2CO$^{\mathrm{top4f}}$ &   --- & 2CO$^{\mathrm{top2f,top4f}}$ & -0.29\\[0.5ex]
\end{tabular}
\end{ruledtabular}
\end{table}
\end{small}

It is also possible to adsorb two CO molecules in the $\sqrt{5}$ surface unit cell, and again we find the adsorption in the two bridge sites  most stable.  Mixing the adsorption sites, i.e. placing e.g. one CO in a bridge site and a second one in a top2f site does always yield less stable structures than the adsorption in like sites (cf. Table~\ref{tab:Eb_sqrt5}).  It also becomes energetically very unfavorable to place more than two CO molecules in the $\sqrt{5}$ surface unit cell as deduced from test calculations with up to 4 CO molecules per surface unit cell.

The adsorption of oxygen in the high-symmetry sites of the $\sqrt{5}$ surface oxide gives similar results.  Again, the bridge site is the most stable site ($E^{\mathrm{bind}}_{\mathrm{O@}\sqrt{5}} = -0.14$\,eV) and the hollow sites are unstable upon relaxation. However, it is now not possible to adsorb any oxygen in  the top2f and top4f sites in the $\sqrt{5}$ unit cell.
If one CO molecule and one O atom are adsorbed simultaneously, likewise the adsorption in two bridge sites is found to be  favored.

Our calculations with one and two adsorbates per $\sqrt{5}$ surface unit cell thus indicate noticeable overall repulsive lateral interactions.  We correspondingly extended our calculations also to more sparse adlayers in $(2 \times 1)$ and $(1 \times 2)$ $\sqrt{5}$ surface unit cells, but did not find significant lateral interactions extending across the $\sqrt{5}$ surface unit cell.  The obtained binding energies were always to within 25\,meV per adsorbate of those calculated for the same adsorption site in a $(1 \times 1)$ cell.

To increase the configuration space of considered overlayers based on the $\sqrt{5}$ surface oxide, we also considered structures, in which the original surface oxide structure is slightly modified.  In a first step, one of the upper hollow site oxygen atoms is removed, and CO and O are again placed in the afore described sites.  Also in this modified structure the bridge site results as  the preferred adsorption site.  If both upper oxygen atoms are removed from the $\sqrt{5}$ surface oxide, we find the structure already  to some extend destabilized.  Due to the lower palladium density in the reconstructed layer (4 Pd atoms on 5 Pd(100) substrate atoms per unit cell) the structure is rather open, and the structural relaxation showed that the Pd atoms can now move quite easily in  lateral direction on the surface.  It is still possible to adsorb O/CO in any of the other adsorption sites, but we found these structures  either unstable upon relaxation or  in general much less stable than the corresponding structures, where the original O hollow vacancy is refilled.

In addition to simply removing the upper oxygen atoms from the $\sqrt{5}$ surface oxide structure and filling additional on-surface adsorption sites, we also substituted the topmost O atoms by CO molecules, and placed additional O atoms and/or CO molecules in top2f, top4f and bridge sites.
In the substituted hollow site the CO binds quite strongly.  However, since the O atoms bind even stronger to this site by 0.6\,eV, O will always preferentially occupy the hollow sites when the two adsorbates compete for this sites.  The situation is reversed for adsorption in the bridge sites.  Regardless of whether O or CO occupies the threefold hollow sites, we always find a stronger binding by at least 0.7\,eV of CO compared to O at these bridge sites.

Even if we leave out the four unstable hollow sites in Fig.~\ref{fig3} there is finally still a huge amount of possible overlayer structures that can be created by combinatorially placing an arbitrary number of O and CO per surface unit cell into any of the available sites.  Since there is only little known about the adsorption on the surface oxide none of these structures can \emph{a priori} be excluded.
Motivated by the strong repulsive interactions seen in our calculations with one or two adsorbates per surface unit cell, we nevertheless discard quite a number of these structures with the criterion that no two adsorbates may sit in directly neighboring sites (i.e. at a distance of less than 1/4 of the length of $\sqrt{5}$ surface unit cell).
This still leaves  92 ``plausible'' structures and DFT calculations were performed for all of them (for more details on the calculated overlayer structures see Ref.~\onlinecite{rogalthesis}).
Out of these  only 55 are stable upon relaxation and are then considered in our constrained ``phase diagram''.

\section{Stability in a constrained equilibrium with an O$_2$ and CO gas phase}
Using the Gibbs free energy of adsorption as defined in Eq.~(\ref{eq:DeltaG}) it is  possible to compare the  stability of all calculated adsorption structures in a constrained equilibrium with an O$_2$ and CO gas phase.  In this approach the surface is considered to be in equilibrium with two separate gas phase reservoirs of O$_2$ and CO, i.e. the formation of CO$_2$ is neglected in the gas phase as well as on the surface~\cite{reuter03b}.  In the gas phase this assumption seems quite justified, since  the direct reaction of O$_2$ and CO is kinetically hindered  by a huge free energy barrier.  On the surface, though, the reaction is actually supposed to take place at the working catalyst.  Here, the assumption of a constrained equilibrium is thus only reasonable as long as the O$_2$ and CO adsorption and desorption events are much more frequent than the reaction, so that the surface can maintain its equilibrium with the two gas phase components.
In other words, the approach amounts to assuming that the kinetics of the on-going catalytic reactions does not change the surface structure and composition.
The resulting constrained surface ``phase diagram'' can thus only provide a first idea of the possibly relevant surface structures, which needs to be scrutinized  explicitly treating the surface kinetics.  On the other hand the constrained equilibrium approach allows to rapidly screen a huge number of (structurally quite different) surface configurations and compare their stability over a wide $(T,p)$-range of possible gas phase conditions.
By using \emph{ab initio} thermodynamics we are thus able to identify those regions in $(T,p)$-space that are likely catalytically active and where we then zoom in with more sophisticated methods.
This second, refining step  explicitly includes the effect of the reaction kinetics on the average surface composition (using kinetic Monte Carlo simulations) and the results of  these simulations, which are restricted to the most relevant surface structural models identified here, are discussed in a second paper.
This will then also bring us into a position to scrutinize the validity of the assumption of the constrained equilibrium approach employed here.

\subsection{Surface Phase Diagram}
\begin{figure*}
\scalebox{0.8}{\includegraphics{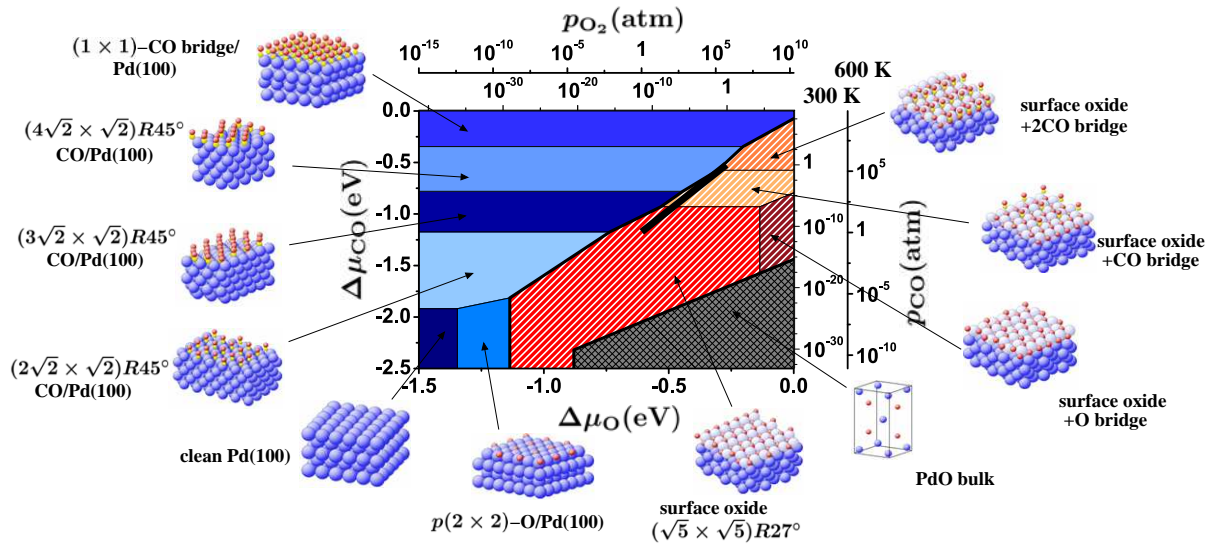}}
\caption{\label{fig4}
(Color online) Surface ``phase diagram'' of the Pd(100) surface in constrained thermodynamic equilibrium with an O$_2$ and CO gas phase.  The atomic structures underlying the various stable (co-)adsorption structures on the metal and the surface oxide, as well as a thick bulk-like oxide film (indicated by the bulk unit cell) are also shown (Pd = large blue spheres, O = small red spheres and C = small yellow spheres). The black bar marks gas phase conditions representative of technological CO oxidation catalysis, i.e.
partial pressures of 1\,atm and temperatures of $300-600$\,K.  The dependence of the chemical potentials of the two gas phases is translated into pressure scales for $T=300$\,K and $T=600$\,K (upper $x$-axes and right $y$-axes).
}
\end{figure*}
The results of the above describe calculations concerning the afore discussed overlayers of oxygen and CO on Pd(100),  also including the (modified) $\sqrt{5}$ surface oxide structure,  are summarized in Fig.~\ref{fig4}, which shows the most stable ``phases'' (i.e. the ones that maximize $\Delta G^{\mathrm{ads}}$ as defined in Eq.~(\ref{eq:DeltaG})) for any given chemical potential of the oxygen and CO gas phase.
The dependence on the chemical potentials has also been converted into more intuitive pressures scales for temperatures of $T=300$\,K and $T=600$\,K in the top two $x$-axes for oxygen and the right two $y$-axes for CO.
As expected from the Gibbs phase rule, stable ``phases'' cover volumes in the 3-dimensional space spanned by $(T, p_{\mathrm{O}_2}, p_{\mathrm{CO}})$, and coexistence regions between two ``phases'' cover areas.  Correspondingly in the 2-dimensional sectional plane shown in Fig.~\ref{fig4} this leads to areas and lines for the two cases, respectively.
Out of the set of 191 tested structures only 11 turn out to be ``phases'' appearing in Fig.~\ref{fig4}.

We will start the discussion of the obtained constrained surface ``phase diagram'' in Fig.~\ref{fig4} in the lower left corner.  Here, both the chemical potential of oxygen and CO are very low, i.e. the oxygen/CO content in the gas phase is insignificant and consequently the clean Pd(100) surface results as the most stable system state.
If we now move along the $x$-axis in Fig.~\ref{fig4} to the right we  reach more and more oxygen-rich conditions, while the CO content in the gas phase is kept low.  The increase in the oxygen chemical potential leads to a stabilization of oxygen containing structures with increasing coverage: First, the $p(2 \times 2)$-O overlayer on Pd(100)\cite{p2x2}, then the $\sqrt{5}$ surface oxide and finally the PdO bulk oxide representing thick bulk-like oxide films on the surface.  This sequence of stable structures was also confirmed by \emph{in situ} surface X-ray diffraction measurements~\cite{lundgren04}.
Interestingly, the $c(2 \times 2)$ structure  observed  under UHV conditions~\cite{orent82,stuve84,chang88a,chang88b,zheng02} does not appear in the surface ``phase diagram''.  Thus we conclude that the $c(2 \times 2)$ structure is most likely a meta-stable state produced by the exposure kinetics.

If  the oxygen content is kept low in the gas phase and the CO content is gradually increased, i.e. moving from the lower left corner along the $y$-axis to the top in Fig.~\ref{fig4},  a series of ordered CO adlayer structures with increasing coverage is stabilized on Pd(100). At first we find the also experimentally observed $(2 \sqrt{2} \times \sqrt{2})R45^{\circ}$, $(3 \sqrt{2} \times \sqrt{2})R45^{\circ}$ and $(4 \sqrt{2} \times \sqrt{2})R45^{\circ}$ structures, and finally for a very high CO content in the gas phase a $(1 \times 1)$ structure with 1\,ML CO in bridge sites.

Starting again in the lower left corner of Fig.~\ref{fig4} we now move along the diagonal, which corresponds to increasing both the oxygen and CO content in the gas phase.  Intuitively, one would expect co-adsorption structures of O and CO on Pd(100)  to become favorable.  But none of the above discussed ordered co-adsorption structures   (cf. Fig.~\ref{fig2})  are found to be a most stable ``phase'' under any gas phase conditions.
This is consistent with the already mentioned experimental findings, that O and CO prefer to form separate domains rather than ordered co-adsorbed overlayers~\cite{stuve84}.  The lower stability of such mixed structures can be explained by the strongly repulsive interactions between adsorbed O and CO, which lead to a significant decrease in binding energies (cf. Table~\ref{tab:Eb_simple}).  However, we can of course not exclude that ordered arrangements with  different periodicities than those considered here would not lead to a lowering in the repulsive interactions.  To take a reasonable part in the ``phase diagram'', though, the binding energies would have to increase by as much as 0.3--0.5\,eV per O atom/CO molecule compared to the now proposed structures. Compared to the pure overlayer structures this would even imply the necessity of attractive interactions between the two species.

For high oxygen and CO content in the gas phase (upper right part of Fig.~\ref{fig4}) we find instead co-adsorbed structures involving the $\sqrt{5}$ surface oxide (two orange-white hatched regions) to become  stable.
These two mixed structures correspond to the surface oxide with one and two CO molecules adsorbed in bridge sites, respectively.  In a small range of very oxygen-rich and intermediate CO gas phase conditions additional O adsorption in the bridge sites of the $\sqrt{5}$ structure leads also to a stable ``phase'' (dark red-white hatched region).

Looking again at the whole surface ``phase diagram'' in Fig.~\ref{fig4} the 11  ``phases'' can be divided into three groups:  on-surface adlayer structures of O or CO on Pd(100), phases involving the $\sqrt{5}$ surface oxide structure, and the  stability region of the bulk oxide as discussed in section~\ref{subsec:bulkoxide}.
Focusing on these three different groups of phases two important conclusions with respect to the relevance of  oxide formation under catalytic reaction conditions can be drawn. First, the formation of a thick, bulk-like oxide (grey cross-hatched area) at the surface
under technologically relevant gas phase conditions of O$_2$ and CO (black bar in Fig.~\ref{fig4}, $p_i \sim 1$\,atm, $T \sim 300-600$\,K) can be ruled out.
This is thus in marked contrast to the much wider stability range of bulk RuO$_2$, which does extend to these conditions~\cite{reuter04a}.
Second, the stability region of the $\sqrt{5}$ surface oxide (hatched area) does extend to such conditions. In fact, it is either this monolayer thin surface oxide or  CO adlayers on Pd(100), which are  neighboring ``phases'' around the catalytically active region in $(T,p)$-space.
As can be seen in Fig.~\ref{fig4} the  catalytically relevant conditions are actually right at the boundary between the $\sqrt{5}$ surface oxide structures and the CO overlayer structures on Pd(100).  Small changes in the Gibbs free energy of adsorption
causing a shift in this boundary could thus well
affect the conclusion as to which is the most stable ``phase'' under these conditions.
Apart from the assumption of a \emph{constrained} thermodynamic equilibrium,
the main uncertainties, which may cause such changes in $\Delta G^{\mathrm{ads}}$ are the approximate DFT total energies and the neglected free energy contributions.
We verified that the uncertainties in the DFT total energies due to the numerical approximations (supercell setup, finite basis set) are not significant in this respect.  What needs to be scrutinized are therefore the uncertainties due to the approximate xc-functional underlying the DFT total energies and the neglected free energy contributions to $\Delta G^{\mathrm{ads}}$.  This will be done in the next two subsections.

\subsection{Evaluating the Gibbs Free Energy}
To calculate the Gibbs free energy of adsorption $\Delta G^{\mathrm{ads}}$ the Gibbs free energies of the different components entering Eq.~(\ref{eq:DeltaG}) have to be evaluated.  The Gibbs free energy
\begin{eqnarray}
\label{eq:G}
G (T,p) = E^{\mathrm{tot}} + F^{\mathrm{vib}} - TS^{\mathrm{conf}} + pV
\end{eqnarray}
comprises contributions from the total energy $E^{\mathrm{tot}}$, the vibrational free energy $F^{\mathrm{vib}}$ including the zero point energy (ZPE), the configurational entropy $S^{\mathrm{conf}}$ and the $pV$-term.
In Eq.~(\ref{eq:DeltaG}) we substituted the Gibbs free energy difference by only  the leading term, the total energy differences, which can be directly obtained from DFT calculations.  We assess the uncertainty introduced by this approximation  by an order of magnitude estimate of the remaining contributions to $\Delta G^{\mathrm{ads}}$.  If this first approximation reveals that the results are significantly influenced by considering all contributions to the Gibbs free energy, the respective terms have to be calculated explicitly.  However, the order of magnitude estimate can be obtained very easily and is thus helpful to decide whether or not it becomes necessary to evaluate the entire Gibbs free energy.

\begin{figure}
\scalebox{0.35}{\includegraphics{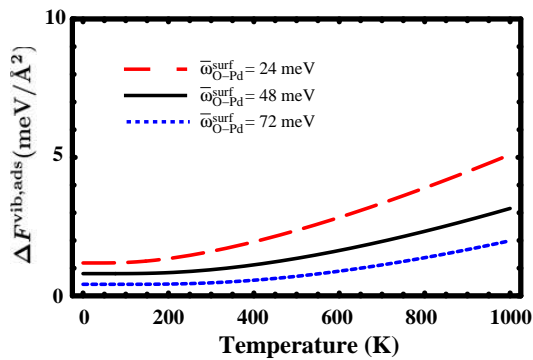}}
\caption{\label{fig5}
(Color online) Vibrational contribution to the Gibbs free energy of adsorption for the $p(2 \times 2)$-O adlayer on Pd(100), within the Einstein approximation using different characteristic frequencies (see ``inset'' and text).
}
\end{figure}

Following the discussion in Refs.~\onlinecite{reuter02} and~\onlinecite{reuter03b} the contributions to the Gibbs free energy of adsorption arising from the $pV$-term and the configurational entropy are rather small for systems like the one presented here.  We will thus only discuss the most crucial approximation, namely the neglect of the vibrational free energy contribution, $\Delta F^{\mathrm{ads, vib}}$ to the Gibbs free energy of adsorption.
In order to estimate the size of   $\Delta F^{\mathrm{ads, vib}}$ we approximate the phonon density of states (DOS) within the Einstein model by one characteristic frequency. Following the approach outlined in Ref.~\onlinecite{reuter02}  the vibrational contribution to the Gibbs free energy of adsorption for the $p(2 \times 2)$-O/Pd(100) structure can then be estimated as
\begin{eqnarray}
\label{eq:Gvib}
\lefteqn{\Delta F^{\mathrm{ads,vib}}(T) \approx} \\ \nonumber
& \approx &
-\frac{1}{A}\left(F^{\mathrm{vib}}(T,\bar{\omega}^{\mathrm{surf}}_{\mathrm{O-Pd}}) - \frac{1}{2}F^{\mathrm{vib,ZPE}}(\bar{\omega}^{\mathrm{gas}}_{\mathrm{O}_2})\right)
\quad ,
\end{eqnarray}
where $F^{\mathrm{vib}}(T,\omega)$ is the frequency dependent function
\begin{eqnarray}
F^{\mathrm{vib}}(T, \omega) =
\frac{1}{2} \hbar \omega + k_{\mathrm{B}} T \ln(1-e^{-\beta \hbar \omega})
\end{eqnarray}
with $\beta = 1/k_{\mathrm{B}} T$.
$\bar{\omega}^{\mathrm{surf}}_{\mathrm{O-Pd}}$ is the characteristic vibrational frequency of an oxygen atom adsorbed in a fourfold hollow site on Pd(100), and $\bar{\omega}^{\mathrm{gas}}_{\mathrm{O}_2}$ is the stretch frequency of the O$_2$ gas phase molecule.  The change in the vibrational contribution of the Pd atoms in the clean and oxygen covered Pd(100) is neglected, so that the vibrational contribution to $\Delta G^{\mathrm{ads}}$ is given by the \emph{difference} in the vibrational energy of the O$_2$ molecule in the gas phase and the oxygen atom on the surface.  For the oxygen molecule only the  ZPE  (not included in the presented DFT total energies) has to be considered, since all remaining contributions to $F^{\mathrm{vib}}$ are already contained in $\Delta \mu_{\mathrm{O}}$ (cf. Eq.~(\ref{eq:DeltaG})).
In Fig.~\ref{fig5} the resulting vibrational contribution $\Delta F^{\mathrm{ads,vib}} (T)$ is shown for a characteristic frequency of $\bar{\omega}^{\mathrm{surf}}_{\mathrm{O-Pd}} = 48$\,meV for the Pd-O stretch frequency of an adsorbed O atom~\cite{nyberg83} and an O-O vibrational frequency of $\bar{\omega}^{\mathrm{gas}}_{\mathrm{O}_2} = 196$\,meV~\cite{herzberg50}.
We find that for the  $p(2 \times 2)$-O/Pd(100) structure the vibrational contribution to $\Delta G^{\mathrm{ads}}$ stays below  $\sim 3$\,meV/\AA$^2$ for temperatures up to $T = 1000$\,K (black-solid line in Fig.~\ref{fig5}).  Since the chosen $\bar{\omega}^{\mathrm{surf}}_{\mathrm{O-Pd}}$ is only a rough guess, we also allow for a $\pm 50$\,\% change of this value, but even then the contribution does not increase considerably (red-dashed, $-50$\,\% and blue-dotted, $+50$\,\%, lines in Fig.~\ref{fig5}).

Similar estimates have been performed for every structure considered for the surface ``phase diagram'' in Fig.~\ref{fig4}.  For CO containing structures $\Delta F^{\mathrm{ads,vib}}$ comprises two different contributions when comparing the gas phase and the adsorbed state, namely the change of the C-O stretch vibration due to the adsorption and an additional Pd-C vibration.
For structures involving the $\sqrt{5}$ surface oxide an additional contribution arises from the change in the vibrational energy between bulk and surface Pd atoms.  The $\sqrt{5}$ structure contains one surface atom less than the corresponding Pd(100) surface per unit cell which also has to be balanced by the bulk reservoir ($\Delta N_{\mathrm{Pd}} = -1$ in Eq.~(\ref{eq:DeltaG})).

With this procedure we find that for all structures $\Delta F^{\mathrm{ads,vib}}$ stays always below 10\,meV/\AA$^2$ for temperatures up to $T = 600$\,K.
The surface ``phase diagram'' discussed in the previous section is not significantly changed, if these estimated, maximum values for the vibrational contribution are included. There are some small shifts in the boundaries between stable phases, but none of the stable structures disappears from and none of the unstable ones appears in the ``phase diagram''.
Furthermore, the boundary between the surface oxide structures and the CO adlayers on Pd(100) is only marginally affected and consequently remains in the vicinity of the catalytic active region in $(T,p)$-space.
We are thus confident that for the here discussed results the approximation of the Gibbs free energy differences by only the total energy terms in Eq.~(\ref{eq:DeltaG}) is justified.

\subsection{Influence of the xc-functional}
\begin{figure}
\scalebox{0.45}{\includegraphics{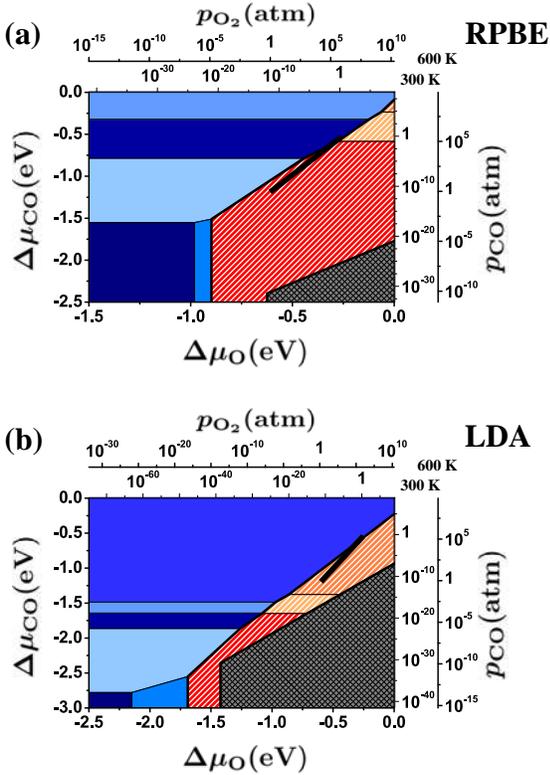}}
\caption{\label{fig6}
(Color online) Surface ``phase diagram'' for the Pd(100) surface in a constrained equilibrium with an O$_2$ and CO gas phase.  The top plot shows the RPBE results and the bottom one the LDA results.  The color coding of the different phases is equivalent to the one used in Fig.~\ref{fig4}. The black bar marks again gas phase conditions representative of technological CO oxidation catalysis, i.e. partial pressures of 1\,atm and temperatures between 300--600\,K.  Note that the shown range of chemical potentials is larger for the LDA results, compared to the RPBE and the range in Fig.~\ref{fig4}.
}
\end{figure}

To obtain an estimate of the uncertainty in Fig.~\ref{fig4} due to the choice of the PBE-GGA xc-functional we reevaluate the constrained surface ``phase diagram'' using the RPBE~\cite{hammer99} and LDA~\cite{perdew92} xc-functionals.
The corresponding surface ``phase diagrams'' are shown in Fig.~\ref{fig6} and are to be compared with Fig.~\ref{fig4}.  The ``phase diagram'' obtained by using the RPBE approximation for the xc-energy (top graph, Fig.~\ref{fig6}(a)) looks in fact very similar to the previously discussed one.  There are some shifts in the actual phase boundaries, and the stability regions of the high coverage phases of oxygen and CO, the $\sqrt{5}+$O$^{\mathrm{br}}$ and $(1\times1)$-CO$^{\mathrm{br}}$/Pd(100) structures, are  shifted outside the shown range of chemical potentials. Yet, the overall topology is fully conserved.
In agreement with the PBE results the bulk oxide is not a  stable ``phase'' anywhere near  ambient gas phase conditions. Due to the smaller heat of formation (cf. Table.~\ref{tab:Eb_gas}), the stability region of the bulk oxide is in fact much further away from this region.  Most importantly, the boundary between the simple overlayer structures on Pd(100)  and the phases based on the $\sqrt{5}$ surface oxide structure (hatched area) is only very little influenced, so that the  most interesting region for oxidation catalysis (black bar)  is again right at the transition between a CO covered Pd(100) surface and phases involving the surface oxide.

Even if the LDA is used as xc-functional (bottom plot, Fig.~\ref{fig6}(b)) the stability region of the bulk oxide does not extend to the catalytically relevant gas phase conditions.
It covers a much larger range compared to the ``phase diagrams'' obtained with the two GGA functionals though. It should also be noted that in the shown LDA ``phase diagram'' the range of O and CO chemical potentials is enlarged to  lower values to include the stability region of the clean metal surface, which in the LDA appears at much lower gas phase concentrations.
This is a consequence of the strong overbinding in the LDA, which stabilizes any adsorbate structure at much lower pressures at the surface.
Nevertheless, the range of technologically relevant catalytic gas phase conditions (black bar) lies again  right at the boundary between the CO covered Pd(100) surface and $\sqrt{5}$ surface oxide structures.

From these results, we conclude that the choice of the xc-functional does have a strong influence on the absolute values of the binding energies, and thereby also on the location of most of the phase boundaries in Fig.~\ref{fig4} and~\ref{fig6}.
However, particularly  the boundary between the CO overlayer structures on Pd(100) and the $\sqrt{5}$ surface oxide is almost unchanged.  The conclusion on the proximity of both phases to the catalytically relevant gas phase conditions seems therefore untouched by the uncertainty due to the approximate xc-functional.

\section{Conclusions}
The stability of the Pd(100) surface has been investigated in a constrained thermodynamic equilibrium with a two component gas phase consisting of O$_2$ and CO.  In this approach the formation of CO$_2$ in the gas phase and at the surface is not considered, such that the effect of the surrounding gas phase on the surface structure and composition is modeled (as a first approximation) through the contact with independent reservoirs representing the reactants.
To establish the actual  surface ``phase diagram'' a large number of different structures with O and CO adsorbed in high symmetry sites on the Pd(100) surface and on the $\sqrt{5}$ surface oxide structure have been considered.  We find that under gas phase conditions of ambient temperatures and pressures, as applied in heterogenous oxidation catalysis, it is either the nanometer thin surface oxide structure or a CO covered Pd(100) surface that is stable, whereas the stability region of the bulk oxide does not extend to these gas phase conditions~\cite{reuter04a}.

To obtain an estimate of the uncertainty introduced by the choice of the specific xc-functional the total energies entering the surface ``phase diagram'' have been calculated  using the PBE, RPBE and LDA.
Comparing the ``phase diagrams'' for these three different xc-functionals, partly dramatic shifts in the positions of the boundaries between the different stable phases  can be observed.  Yet, it can also be seen that the position of the boundary between the CO covered Pd(100) phases and the $\sqrt{5}$ surface oxide structures is in fact little affected.  For all three xc-functionals, the catalytically active region is very close to this boundary, but just still within the stability range of the surface oxide.
We also verified that this finding is not affected by the numerical uncertainties and the free energy contributions neglected in our approach.

These results suggest that the only monolayer thin surface oxide structure might indeed be a  relevant structure for the CO oxidation reaction on Pd(100) at technological relevant pressures. It has to be considered, though, that there are still two notable approximations in the constrained atomistic thermodynamics
approach, as it has been applied here.
First, configurational entropy is not included, and  at finite temperatures a first effect  would be to smear out the phase boundaries in the ``phase diagram'' and create coexistence regions, where the energetically low lying configurations are mixed according to the law of mass action.
At the phase boundary relevant for CO oxidation catalysis, this could e.g. lead to the formation of coexisting domains of CO covered Pd(100) and surface oxide patches.
Second, the neglected kinetics of the on-going catalytic CO$_2$ formation might significantly change the stability range of the different phases as obtained in the constrained equilibrium approach.
A possible coexistence between patches of metallic Pd(100) covered by the reactants, and patches of the surface oxide (possibly also with adsorbed CO) could then even under steady-state conditions go hand in hand with a continuous formation and decomposition of the oxidic phase.  Such oscillations in the morphology of the catalyst surface could  again significantly  influence the catalytic function of this surface.

Based on the insight described in this paper we thus identify either the $\sqrt{5}$ surface oxide or CO covered Pd(100) as surface structures most relevant under catalytically interesting gas phase conditions.
It is therefore specifically the stability of the $\sqrt{5}$ surface oxide against CO induced decomposition, which needs to be scrutinized to assess the active state of the surface under reaction conditions.
In a second paper following the present one, we correspondingly extend our study to investigate this stability.
Using kinetic Monte Carlo simulations we then explicitly scrutinize the two major uncertainties of the present approach, namely  the effect of configurational disorder and the kinetic effects due to the on-going reactions.

\section{Acknowledgement}
We acknowledge detailed discussions with M. Todorova and Y. Zhang.
The EU is acknowledged for financial support under contract NMP3-CT-2003-505670 (NANO$_2$), and the DFG for support within the priority program SPP1091.


\begin{thebibliography}{99}
\bibitem{p2x2}
We note that in our earlier calculations~\cite{todorova03,reuter04a} the $p(2 \times 2)$  overlayer did not appear as stable structure.  The discrepancy between these earlier results and the here presented data is partly due to an insufficent convergence of the earlier DFT calculations, as well as to a wrong treatment of the difference in the Pd atom density ($\Delta N_{\mathrm{Pd}} \neq 0$) between the Pd(100) surface and the reconstructed $\sqrt{5}$ surface oxide, when calculating the Gibbs free energy of adsorption.
\end{thebibliography}
\end{document}